\documentstyle[preprint,eqsecnum,aps]{revtex}
\begin{document}
\draft
\preprint{\parbox{4cm}{\flushright CALT-68-2008\\ MZ-TH/95-20}}
\title{More Conservation Laws \\and Sum Rules in the Heavy Quark Limit}
\author{Chi-Keung Chow}
\address{California Institute of Technology, Pasadena, CA 91125
\footnote{Present address:  Newman Laboratory of Nuclear Studies,
Cornell University, Ithaca, NY 14853.}.}
\author{Dan Pirjol}
\address{Johannes Gutenberg-Universit\"at, Institut f\"ur Physik (THEP),\\
Staudingerweg 7, D-55099 Mainz, Germany.}
\date{\today}
\maketitle
\begin{abstract}
This is the continuation of a previous article, in which the Bjorken and
Voloshin sum rules were interpreted as statements of conservation of
probability and energy.
Here the formalism is extended to higher moments of the Hamiltonian operator.
{}From the conservation of the second moment of the Hamiltonian operator one
can
derive a sum rule which, in the small velocity limit, reduces to the
Bigi--Grozin--Shifman--Uraltsev--Vainshtein (BGSUV) sum rule.
On the other hand, the conservation of the third moment of the Hamiltonian
operator gives a new
sum rule, which is related to the matrix element of the heavy quark
counterpart of the Darwin term in atomic physics. The general case of
the higher order moments is also discussed.
\end{abstract}
\narrowtext
\section{Introduction}
In a previous article \cite{1}, the relationship between various conservation
laws and various sum rules has been explored.
In the heavy quark limit, the hadronic resonances appear as eigenstates of the
light degrees of freedom under the static color field of the heavy quark.
Under a $b\rightarrow c$ decay, the Isgur--Wise form factors $\varphi_{n'}(w)$
are defined to be, up to a kinematic factor, the overlap of the initial ground
state brown muck $|0\rangle$ moving with velocity $v$ and the final (ground
state or excited) brown muck $|n'\rangle$ with velocity $v'$,
\begin{equation}
\varphi_{n'}(w)=\langle n'|0\rangle,
\end{equation}
with $w=v\cdot v'$.
By the completeness of the set of eigenstates, we have the master sum rule
\begin{equation}
\sum_{n'} \langle 0|{\bf X}|n'\rangle \langle n'|0\rangle
= \langle 0|{\bf X}|0\rangle,
\label{msr}
\end{equation}
where ${\bf X}$ is an arbitrary operator.
The case of interest is when ${\bf X}$ is conserved, i.e., $H'$ and ${\bf X}$
commute and the $|n'\rangle$'s are eigenstates of ${\bf X}$.
Then the left hand side of Eq. (\ref{msr}) is just the sum of the squares of
the Isgur--Wise form factors $\varphi(w)$ weighted by their respective
eigenvalues of ${\bf X}$.
For ${\bf X} = \openone$, the identity matrix, this master sum rule is
equivalent to the well-known Bjorken sum rule \cite{2,3,4},
\begin{equation}
\sum_{n'} |\varphi_{n'}(w)|^2  = 1.
\label{bsr}
\end{equation}
On the other hand, with ${\bf X} = H'$, the hamiltonian operator in the $v'$
frame, Eq. (\ref{msr}) can be reduced to the Voloshin sum rule \cite{5} (which
will be discussed below).
Lastly, a new sum rule can be obtained by choosing ${\bf X}=P'$, the parity
operator in the $v'$ frame.
This new ``parity sum rule'' can lead to model independent lower bounds of
$\varphi_{0'}(w)$, the ground state Isgur--Wise form factor.

In this paper, we are going to consider the case ${\bf X}=H'^k$ with $k\geq2$.
When ${\bf X}=H'^2$, the sum rule obtained is closely related to the
Bigi--Grozin--Shifman--Uraltsev--Vainshtein (BGSUV) sum rule derived in
Ref. \cite{6}.
When ${\bf X} = H'^3$, a new sum rule, relating the Isgur--Wise form factors
and the matrix element of the heavy quark counterpart of the Darwin term in
atomic physics, is obtained. The general case of ${\bf X} = H'^k$ with
$k\geq 4$ is briefly discussed.

\section{Sum Rules On $H'^2$}
By considering the operator product expansion of heavy quark currents in the
context of QCD sum rules, the authors of Ref. \cite{6} have been able to obtain
several sum rules relating the Isgur--Wise form factors in the small velocity
limit.
Besides the Bjorken and Voloshin sum rules, they obtained the new BGSUV sum
rule,
\begin{equation}
\sum_{n'} |\varphi_{n'}(w)|^2 E_{n'}^2 = \textstyle{1\over3} \mu_\pi^2
(\vec v-\vec v')^2,
\label{H2}
\end{equation}
where $E_{n'}=m_{X^{n'}_b}-m_B$ is the mass difference over the ground state
meson, and
\begin{equation}\label{fff}
\mu_\pi^2=\langle B|\bar b(i\vec D)^2 b|B\rangle.
\end{equation}
(Notice that the present notations are different from those in Ref. \cite{5}.
In particular the normalizations of the meson states differ by a factor of
$\sqrt{2M_B}$.)

Physically $\mu_\pi^2$ denotes the square of the {\it heavy quark} momentum
inside the $B$ meson, which is also the square of the {\it brown muck} momentum
in the hadron rest frame.
\begin{equation}\label{sss}
\mu_\pi^2 = \langle 0|\vec p\,^2|0 \rangle.
\end{equation}
Evidently this sum rule expresses conservation of the second moment of the
energy operator, and we will see that this sum rule indeed can be reproduced
as a special case of our master sum rule.

Putting ${\bf X}=H'^2$ in the master sum rule, we have
\begin{equation}
\sum_{n'} |\varphi_{n'}(w)|^2 \Delta m_{n'}^2 = \langle 0|H'^2|0\rangle,
\label{h2}
\end{equation}
with $\Delta m_{n'} = m_{X^{n'}_b}-m_{b}$ as the mass of the ``brown muck''.
To see how the quantity $\langle 0|H'^2|0\rangle$ can be evaluated, it is
worthwhile to review briefly the derivation of the Voloshin sum rule \cite{1}.
The Voloshin sum rule is obtained by putting ${\bf X}=H'$ in Eq.
(\ref{msr}).
\begin{equation}
\sum_{n'}\Delta m_{n'}|\varphi_{n'}(w)|^2 = \langle 0|H'|0\rangle.
\label{h}
\end{equation}
To calculate the right-hand side, note that $|0\rangle$ and $|0'\rangle$
are related by a Lorentz transformation $L$ which boosts from the
$v'$ frame to the $v$ frame.
\begin{equation}
|0\rangle = L|0'\rangle.
\end{equation}
Then
\begin{equation}
\langle 0|H'|0\rangle = \langle 0'|L^{-1}H'L|0'\rangle.
\end{equation}
The Lorentz transformation dictates that the hamiltonian is transformed as the
time component of a four vector.
\begin{equation}
L^{-1}H'L = w(H' + (\vec v-\vec v\,')\cdot\vec p\,').
\end{equation}
Since the ground state $|0'\rangle$ is at rest in the $v'$ frame,
$\langle 0'|\vec p\,'|0'\rangle = 0$.
On the other hand, $\langle 0'|H'|0'\rangle = \Lambda = m_B-m_b$ is the mass of
the brown muck.
Hence,
\begin{equation}
\langle 0|H'|0\rangle = w \langle 0'|H'|0' \rangle
+ w(\vec v-\vec v\,')\cdot\langle 0'|\vec p\,'|0'\rangle = w\Lambda,
\end{equation}
as conjectured in Ref. \cite{1}.
Together with Eq. (\ref{h}), the Voloshin sum rule is reproduced:
\begin{equation}
\sum_{n'}\Delta m_{n'}|\varphi_{n'}(w)|^2 = w\Lambda.
\label{vsr}
\end{equation}

The argument above can be generalized to evaluate $\langle 0|H'^2|0\rangle$.
\begin{equation}\label{symmetric}
\langle 0|H'^2|0\rangle = w^2 \langle 0'|H'^2|0' \rangle + w^2 (v-v')_i
(v-v')_j \langle 0'|p'_i p'_j|0' \rangle.
\end{equation}
with the terms proportional to $\langle 0'|\vec p\,'|0'\rangle$ vanishing.
By the relation
\begin{equation}
\langle 0'|p'_i p'_j|0' \rangle = \textstyle{1\over3} \langle 0'|\vec p\,'^2|
0' \rangle \delta_{ij} = \textstyle{1\over3} \mu_\pi^2 \delta_{ij},
\label{sym}
\end{equation}
we end up with
\begin{eqnarray}
\sum_{n'} |\varphi_{n'}(w)|^2 \Delta m_{n'}^2 &=& \langle 0|H'^2|0\rangle
\nonumber\\ &=& w^2 \Lambda^2 + \textstyle{1\over3} w^2 \mu_\pi^2 (\vec v -
\vec v\,')^2.
\label{2sr}
\end{eqnarray}
(Strictly speaking, the matrix element (\ref{sym}) contains in general also
an antisymmetric part in the indices $i,j$. However, for the purpose of
deriving the sum rule (\ref{2sr}) this part can be neglected, as it
vanishes under multiplication with $(v-v')_i(v-v')_j$ in Eq.
(\ref{symmetric}).)

Using Eqs. (\ref{bsr}), (\ref{vsr}), (\ref{2sr}), and the relation $E_{n'} =
\Delta m_{n'} - \Lambda$, the left-hand side of Eq. (\ref{H2}) can be
evaluated.
\begin{eqnarray}
\sum_{n'} |\varphi_{n'}(w)|^2 E_{n'}^2 &=& \sum_{n'} |\varphi_{n'}(w)|^2
(\Lambda^2 - 2\Lambda\Delta m_{n'} + \Delta m_{n'}^2) \nonumber\\
&=& \Lambda^2 - 2w \Lambda^2 + w^2 \Lambda^2
+ \textstyle{1\over3} w^2 \mu_\pi^2 (\vec v - \vec v')^2\nonumber\\
&=& \Lambda^2 (w-1)^2 + \textstyle{1\over3} w^2 \mu_\pi^2 (\vec v - \vec v')^2.
\label{csr}
\end{eqnarray}
So in the small velocity limit, i.e., $w\simeq1$,
\begin{equation}
\sum_{n'} |\varphi_{n'}(w)|^2 E_{n'}^2
= \textstyle{1\over3} \mu_\pi^2 (\vec v - \vec v')^2,
\end{equation}
and the BGSUV sum rule is reproduced.

Noting that $(\vec v - \vec v')^2 = w^2 - 1$, Eq. (\ref{csr}) can be formally
expanded about $w=1$ as
\begin{equation}
\sum_{n'} |\varphi_{n'}(w)|^2 E_{n'}^2 = \textstyle{2\over3} \mu_\pi^2 (w-1) +
\dots.
\label{expn}
\end{equation}
On the other hand, we know that in general, for a state with orbital angular
momentum $l>0$, $|\varphi_{n'}(w)|^2 \sim (w-1)^l$.
For $l=0$,  $|\varphi_{n'}(w)|^2 \sim (w-1)^0$ for the ground state and $\sim
(w-1)^2$ for excited states.
This determines the behavior of the $\varphi$'s near the point of zero recoil.

The zeroth order expansion of Eq. (\ref{expn}) about the point of zero recoil
is the trivial identity $0=0$, while the first non-trivial term is
\begin{equation}
\sum_{l=1} |\varphi_{n'}(w)|^2 E_{n'}^2 = \textstyle{2\over3} \mu_\pi^2 (w-1)
= \textstyle{1\over3} \mu_\pi^2 (v-v')^2,
\end{equation}
where $v$ and $v'$ are the {\it four} velocities of the initial and final
heavy quarks.
Then the above modification of the BGSUV sum rule is saturated by the $P$-wave
states.
(In contrast, the original BGSUV sum rule runs over all excited states.)

In the meson sector, this modified BGSUV sum rule becomes
\begin{equation}\label{BGSUV}
\sum_q 2\left(E_{1/2}^{(q)}\right)^2 |\tau_{1/2}^{(q)}(1)|^2 + \sum_r 4\left(
E_{3/2}^{(r)}\right)^2 |\tau_{3/2}^{(r)}(1)|^2 = \textstyle{2\over3} \mu_\pi^2,
\end{equation}
where the $P$-wave Isgur--Wise form factors $\tau(w)$'s are defined in Ref.
\cite{4}.
Although we have derived this modified BGSUV sum rule in the context of $B$
decays, the same discussion holds for $\Lambda_b$ decays as well, for which the
sum rule reads
\begin{equation}
\sum_q 2\left(E_1^{(q)}\right)^2 |\sigma^{(q)}(1)|^2 = \textstyle{2\over3}
\bar\mu_\pi^2,
\end{equation}
where $\bar\mu_\pi^2$ is defined in analogy to $\mu_\pi^2$
\begin{equation}
\bar\mu_\pi^2=\langle\Lambda_b|\bar b(i\vec D)^2 b|\Lambda_b\rangle,
\end{equation}
and the $P$-wave Isgur--Wise form factors $\sigma(w)$'s are defined in Ref.
\cite{3}.

\section{Sum Rule on $H'^3$}
In this section we will derive a sum rule from the conservation of the
third moment of the hamiltonian.
Bearing in mind that $H$ and $p$ are operators which in general do not commute,
putting ${\bf X} = H'^3$ in the master sum rule gives
\widetext
\begin{eqnarray}
\sum_{n'}|\varphi_{n'}(w)|^2\Delta m_{n'}^3&=&\langle0|H'^3|0\rangle\nonumber\\
&=&w^3 \langle0'|(H'+(\vec v-\vec v\,')\cdot\vec p\,')(H'+(\vec v-\vec v\,')
\cdot\vec p\,')(H'+(\vec v-\vec v\,')\cdot\vec p\,')|0'\rangle\nonumber\\
&=&w^3 \langle0'|H'^3|0'\rangle + w^3 (\vec v - \vec v\,')_i (\vec v - \vec v\,
')_j\langle 0'|(H' p'_i p'_j + p'_i H' p'_j + p'_i p'_j
H')|0'\rangle\nonumber\\
&=&w^3 \Lambda^3 + w^3 \Lambda \mu_\pi^2 (\vec v - \vec v\,')^2
+ \textstyle{1\over3} w^3 \mu_\chi^3 (\vec v - \vec v\,')^2,
\end{eqnarray}
where
\begin{equation}\label{muchi}
\mu_\chi^3 = \langle 0|(p_i H p_i - p_i p_i H)|0\rangle
= \langle0|p_i [H, p_i]|0\rangle.
\end{equation}
Then
\begin{eqnarray}
\sum_{n'} |\varphi_{n'}(w)|^2 E_{n'}^3 &=& \sum_{n'} |\varphi_{n'}(w)|^2
(- \Lambda^3 + 3 \Lambda^2\Delta m_{n'} - 3\Lambda\Delta m_{n'}^2 + \Delta
m_{n'}^3) \nonumber\\
&=& - \Lambda^3 + 3\Lambda^3w - 3\Lambda(w^2\Lambda^2 + \textstyle{1\over3} w^2
\mu_\pi^2 (\vec v - \vec v')^2)\nonumber\\
& &\qquad\qquad + (w^3 \Lambda^3 + w^3 \Lambda \mu_\pi^2 (\vec
v - \vec v')^2 + \textstyle{1\over3} w^3 \mu_\chi^3 (\vec v - \vec v')^2)
\nonumber\\
&=& \Lambda^3 (w-1)^3 + w^2 (w-1) \Lambda \mu_\pi^2 (\vec v - \vec v')^2
+ \textstyle{1\over3} w^3 \mu_\chi^3 (\vec v - \vec v')^2.
\end{eqnarray}
\narrowtext
The first, second and third terms are of order $(w-1)^3$, $(w-1)^2$ and
$(w-1)^1$ respectively.
As a result, when expanded about the point of zero recoil, we get
\begin{equation}\label{sumrule}
\sum_{l=1} |\varphi_{n'}(w)|^2 E_{n'}^3 = \textstyle{2\over3} \mu_\chi^3
(w-1).
\end{equation}

In the heavy mass limit, the commutators of the momentum and hamiltonian
brown muck appearing in this matrix element can be written only in terms of
heavy quark and gluon fields. The result for $\mu_\chi^3$ reads
(details of the derivation can be found in the Appendix)
\begin{equation}\label{final}
\mu_\chi^3 = \frac12 \langle B|\bar Q\gamma^0 g(D^\mu F_{0\mu}^a)t^aQ
|B\rangle\,.
\end{equation}

A sum rule similar to Eq. (\ref{sumrule}) is obtained in the framework of
\cite{6}, from considering the third momentum of the hadronic tensor (in the
terminology of \cite{6}, the ``fourth sum rule''). We have explicitly checked
that the two methods lead to identical results.

   The matrix element $\mu_\chi^3$ (\ref{muchi}) has a special significance
for heavy hadron physics. The dimension-6 operator in (\ref{final})
appears in the heavy quark effective theory lagrangian at order $1/m^2$.
In nonrelativistic language (in the rest frame of the heavy hadron),
$\mu_\chi^3$ is the expectation value of the Darwin term in the Pauli equation
satisfied by the heavy quark field $Q$.
A similar quantity has been defined as $\rho_D^3$
in Eq.~(27) of \cite{Big} where its contribution to the mass of a heavy
hadron has been computed (see also \cite{Mannel}). Also, $\mu_\chi^3$ is
directly connected to the
third moment of the distribution function $F(x)$ for inclusive semileptonic
decays of heavy hadrons into final massless quarks.
Its value determines the asymmetry of $F(x)$ between positive and negative
values of the scaling variable $x$. Finally, a quantity related to
$\mu_\chi^3$ is needed when computing mass corrections of order $1/m^3$ to
the total semileptonic width of a heavy hadron \cite{BDS}.

   Usually the matrix element (\ref{final}) is evaluated with the help of
the factorization approximation \cite{Internal,Mannel}. For this the equation
of motion of the
gluon field is used $D^\mu F_{0\mu}^a = g\sum_q \bar q\gamma_0t^aq$, where
the sum runs over all light quarks. Then the Fierz identity is applied
followed by the insertion of the vacuum state. The result reads
\begin{equation}
\mu_\chi^3 = \frac{8\pi}{9}\alpha_s f_B^2m_B\,,
\end{equation}
where the case of a B meson has been considered.
To turn this into a numerical prediction, a scale must be chosen at which
the coupling $\alpha_s$ is to be evaluated. We employ for this a low scale of
the order $\mu=0.5$ GeV, where $\alpha_s=0.31$ \cite{BDS}. For $f_B$ we use
an average value of the lattice QCD results $f_B=175\pm 25$ MeV \cite{fB}.
Together with the B meson mass $m_B = 5.28$ GeV these values give the
following estimate
\begin{equation}\label{factnumber}
\mu_\chi^3 = 0.140\pm 0.043 \mbox{ GeV}^3\,.
\end{equation}

   On the other hand, the sum rule (\ref{sumrule}) allows a direct
evaluation of this quantity, in terms of the contributions of the excited
states. Compared with the factorization approach, this method has the
advantage of being free of uncertainties connected with the value of the
factorization scale. Therefore it provides us with an independent test
of the vacuum insertion approximation for hadrons containing one heavy
quark.

   Expressed in terms of the contributions of the P-wave mesons, the sum
rule (\ref{sumrule}) takes a similar form to (\ref{BGSUV}):
\begin{equation}\label{srule}
 2E_{1/2}^3 |\tau_{1/2}(1)|^2 + 4 E_{3/2}^3 |\tau_{3/2}(1)|^2
 = \textstyle{2\over3}\mu_\chi^3\,.
\end{equation}
We have neglected here the contributions from the continuum and the higher
radial excitations.
%We will return to this point later on.
The contributions on the left-hand side can be written \cite{6}, via the
Bjorken sum rule, in terms of the slope $\rho$ of the Isgur-Wise function at
the equal-velocity point $v=v'$:
\begin{equation}
\xi(w) = 1-\rho^2 (w-1) + \cdots\,.
\end{equation}
One has
\begin{equation}\label{slope}
\rho^2 = \textstyle{1\over4} + |\tau_{1/2}(1)|^2 +  2  |\tau_{3/2}(1)|^2\,,
\end{equation}
where again we have assumed saturation by the lowest-lying P-wave states
only. Combining (\ref{srule}) and (\ref{slope}), the following bounds are
obtained for $\mu_\chi^3$
\begin{equation}\label{bounds}
3E_{1/2}^3\left(\rho^2-\textstyle{1\over4}\right) \leq \mu_\chi^3 \leq
3E_{3/2}^3\left(\rho^2-\textstyle{1\over4}\right)\,.
\end{equation}
The excitation energy of the $s_\ell^{\pi_\ell}={3\over2}^+$ P-wave multiplet
can be directly extracted from experiment. In the charm sector the masses
the two members of the doublet are \cite{PDG} $m_{D_1}=2423\pm 3$ MeV and
$m_{D_2^*}=2464\pm 4$ MeV (average values over the isospin doublet).
These are expected to have arisen from a degenerated multiplet in the
infinite mass limit, with a mass $m_{P_{3/2}}=1/8(3m_{D_1}+5m_{D_2^*})=
2449\pm 4$ MeV. The average mass of the corresponding S-wave charmed states
is $m_S=1/4(m_D+3m_{D^*})$. Using $m_D=1867$ MeV and $m_D=2008$ MeV
gives $m_S=1973$ MeV, from which the excitation energy $E_{3/2}=476\pm 4$
MeV is obtained.

  Recently, the discovery of the $s_\ell^{\pi_\ell}={3\over2}^+$ family of
P-wave bottom mesons has been announced \cite{B**1,B**2}. The first of these
collaborations quotes the following mass
values: $m_{B_1}=5725$ MeV and $m_{B_2^*}=5737$ MeV, which combine to
an average $m_{P_{3/2}}=5732.5$ MeV. The average mass of the S-wave bottom
mesons is $m_S=5279$ MeV, which gives an excitation energy $E_{3/2}=453.5$
MeV. The difference from the charmed mesons' case can be attributed to
spin-independent corrections proportional to $1/m_{c,b}$. We will use in
the following this
latter number, which is presumably closer to the static value of this
quantity in the infinite mass limit.

   None of the $s_\ell^{\pi_\ell}={1\over2}^+$ P-wave states has been observed
and the only information we have about them comes from model calculations
\cite{model1,model2}. Their masses in the charm sector are expected to lie
in the
range $m_{P_{1/2}}=2300-2400$ MeV, from which an excitation energy of
about $E_{1/2}=380\pm 50$ MeV can be extracted. In analogy with the
preceding case one expects this value to decrease slightly when the mass
of the heavy quark is increased. Thus we will use in the following
$E_{1/2}=360\pm 50$ MeV.

  The other parameter entering (\ref{bounds}) is the slope of the
Isgur-Wise function at $w=1$, which has been measured by the CLEO
collaboration, with the result \cite{CLEO1,CLEO2} $\rho^2=0.84\pm 0.12\pm
0.08$.
Combining these parameters, the bounds (\ref{bounds}) read
\begin{equation}\label{number}
0.083^{+0.081}_{-0.048} \mbox{ GeV}^3 \leq \mu_\chi^3 \leq
0.165^{+0.050}_{-0.056} \mbox{ GeV}^3\,.
\end{equation}

   Comparing with (\ref{factnumber}) one can see that the factorization
approximation gives a rather good estimate of the matrix element (\ref{final}).
Despite the large errors on the lower limit in (\ref{number}), it seems
justified to conclude that, at the very least, the factorization
approximation provides us with the correct sign of this
quantity\footnote{For the purpose of comparison, we quote also the result
of a calculation based on the ISGW model \cite{4} which gives $\tau_{1/2}
(1)=\tau_{3/2}(1)=0.310$. Together with the excitation energies given in
the text, one obtains from (\ref{srule}) $\mu_\chi^3=0.067$ GeV$^3$.
The slope of the Isgur-Wise function comes out equal to 0.538, well under the
CLEO result \cite{CLEO1,CLEO2}, which seems to imply that $\mu_\chi^3$ is also
underestimated.}.

\section{Higher moments of $H'$}
Sum rules similar to the ones discussed in the preceding sections can be
derived for even higher moments of the Hamiltonian operator.
The general form for these sum rules can be obtained by using
${\bf X}=(H'-\Lambda)^k$ in the master sum rule (\ref{msr}) with
$k\geq 4$. This gives
\begin{equation}
\sum_{n'} |\varphi_{n'}(w)|^2 E_{n'}^k = \langle 0|(H'-\Lambda)^k
|0\rangle = \langle 0'|(wH'+w(\vec v-\vec v\,')\cdot\vec p\,' -\Lambda)^k|
0'\rangle\,.
\end{equation}
The matrix element on the right-hand side can be expanded in powers of
$\vec v-\vec v\,'$ and only the terms of even order give a nonvanishing
contribution. Of these, only those quadratic in $\vec v-\vec v\,'$ will
contribute to order $(w-1)$ (since $(\vec v-\vec v\,')^2=w^2-1$).
One obtains in this way
\begin{eqnarray}
\sum_{n'} |\varphi_{n'}(w)|^2 E_{n'}^k &=&
w^2(\vec v-\vec v\,')_i(\vec v-\vec v\,')_j
\langle 0'|p'_i(wH' -\Lambda)^{k-2}p'_j|0'\rangle + {\cal O}((w-1)^2)\\
&=& \frac13 w^2(w^2-1)\langle 0'|p'_i(wH' -\Lambda)^{k-2}p'_i|0'\rangle
+ {\cal O}((w-1)^2)\nonumber\\
&=& \frac23(w-1)\langle 0'|p'_i(H' -\Lambda)^{k-2}p'_i|0'\rangle +
{\cal O}((w-1)^2)\nonumber\,.
\end{eqnarray}
The matrix element on the right-hand side can be written as (removing the
primes for clarity)
\begin{eqnarray}\label{muk}
\mu_k^{k} &=& \langle 0|p_i(H -\Lambda)^{k-2}p_i|0\rangle =
\langle 0|[p_i,\,H](H -\Lambda)^{k-4}[H,\,p_i]|0\rangle\\
&=& \langle 0|[p_i,\,H]
\overbrace{[H,\,\cdots}^{k-4},\,[H,\,p_i]]\cdots ]
|0\rangle\nonumber\,.
\end{eqnarray}
For $k=2,3$ the corresponding matrix elements have been denoted as
$\mu_\pi^2$ and $\mu_\chi^3$, respectively.

  The multiple commutator in (\ref{muk}) can be computed explicitly with
the help of (\ref{basic}). For example,
\begin{equation}
[H,\,[H,\,p_i]] = [H_t-H_h,\,[H,\,p_i]] = -i\frac{\mbox{d}}{\mbox{d}t}
[H,\,p_i] - [H_h,\,[H,\,p_i]]\,,
\end{equation}
where the notations are explained in the Appendix. The first term is equal
to
\begin{equation}\label{timed}
-i\frac{\mbox{d}}{\mbox{d}t}[H,\,p_i] = \int\mbox{d}\vec x\,
\left\{Q^\dagger (-i\stackrel{\leftarrow}{D_0}) igF_{0i}Q +
Q^\dagger igF_{0i}(-i\vec D_0)Q + Q^\dagger g(D_0F_{0i})Q\right\}\,,
\end{equation}
whereas the second one can be explicitly evaluated with the help of
(\ref{Hh}). By making use of the equation of motion (\ref{eqmotion})
one finds that the latter just cancels the first two terms in (\ref{timed})
so that one obtains
\begin{equation}
[H,\,[H,\,p_i]] = \int\mbox{d}\vec x\,Q^\dagger g(D_0F_{0i})Q\,,
\end{equation}
or, more generally,
\begin{equation}
\overbrace{[H,\,\cdots,\,}^{k-4}[H,\,p_i]\cdots] =
(-i)^{k-1}\int\mbox{d}\vec x\,Q^\dagger g(D_0^{k-4}F_{0i})Q\,.
\end{equation}

   Finally, one has for $\mu_k^k$ (\ref{muk})
\begin{eqnarray}\label{mukfinal}
\mu_k^{k} &=& -i\int\mbox{d}\vec x\,Q^\dagger gF_{0i}Q\,\cdot
(-i)^{k-1}\int\mbox{d}\vec y\,Q^\dagger g(D_0^{k-4}F_{0i})Q\\
&=& (-i)^k\int\mbox{d}\vec x\,Q^\dagger g^2F_{0i}(D_0^{k-4}F_{0i})Q\,.
\nonumber
\end{eqnarray}
In the last step we have made use of the fact that the quark $Q$ is
infinitely heavy and thus heavy quark pair creation is rigorously
forbidden. This implies that a first-quantized (quantum-mechanical)
description for the heavy quark is exact in this limit, with operators
satisfying simple multiplication rules. The final result in (\ref{mukfinal})
has been translated back in a second-quantized language\footnote{A similar
argument has been used implicitly in the derivation of the BGSUV sum rule
in section II when identifying the operator $p_ip_i$ in
(\ref{sss}) with $\bar b(i\vec D)^2b$ in (\ref{fff}).}.

   The general form of our sum rule is thus
\begin{equation}
\sum_{l=1} |\varphi_{n'}(w)|^2 E_{n'}^k = \textstyle{2\over3} (\mu_k)^k
(w-1),\label{general}
\end{equation}
with only P-wave states contributing on the left-hand side.
The matrix element appearing in this sum rule involves (for $k\geq 4$) the
chromoelectric field and its time derivatives. The same quantities have been
shown in \cite{Internal} to play an important role in the inclusive decays
of heavy hadrons to states containing a final massive quark. More
precisely, the matrix element $\mu_k$ is connected to the $(k-2)^{th}$ moment
of the so-called ``temporal'' distribution function for these decays.
Therefore the most important application of the sum rules (\ref{general})
can be expected to be constraining the values of these matrix elements and
thereby giving useful information about the temporal distribution function.
Also, the operators whose matrix elements define $\mu_k$ might appear in
physical applications as power corrections to the infinite quark mass limit.
However, from dimensional arguments they are expected to appear suppressed
by a factor of $1/M_Q^{k-1}$ and hence be unimportant when $k$ is large.

\acknowledgments
C.K.C. would like to thank Mark Wise for valuable discussions.
His work was supported by the U.S. Dept. of Energy under Grant No.
DE-FG03-92-ER 40701.
D.P. acknowledges a grant from the Deutsche Forschungsgemeinschaft (DPG).

\appendix
\section*{}

 This appendix contains the technical details of the computation
of the matrix element of brown muck operators which appears in the
definition of $\mu_\chi^3$ (\ref{muchi}). Our final aim is to express it
only in terms of matrix elements of the heavy quark and gluon fields.
We begin by defining the momentum and hamiltonian operators for the total
system (brown muck plus heavy quark):
\begin{eqnarray}\label{pt}
\vec p_t &=& \vec p + \vec p_h\\
H_t &=& H + H_h\,.\label{Ht}
\end{eqnarray}
We will only need in the following the explicit form of the heavy quark
operators:
\begin{eqnarray}
\vec p_h &=& \int\mbox{d}\vec x\,Q^\dagger (-i\nabla - g\vec A^at^a)Q\\
H_h &=& \int\mbox{d}\vec x\,Q^\dagger [\vec\alpha\cdot (-i\nabla -
g\vec A^at^a) +\beta m]Q\,.\label{Hh}
\end{eqnarray}
$Q$ is the heavy quark field operator and $t^a$ are the generators of the
SU$_c$(3) gauge group, satisfying the commutation relations $[t^a,\,t^b]
=if_{abc}t^c$. The equation of motion satisfied by $Q$ is
\begin{equation}\label{eqmotion}
iD_0 Q = [\vec\alpha\cdot (-i\nabla - g\vec A^at^a) +\beta m] Q
\end{equation}
with $D_\mu=\partial_\mu+igA_\mu^a t^a$.

  Let us consider the commutator between the heavy quark hamiltonian and
the heavy quark momentum. It can be written as
\begin{eqnarray}
[H_h,\, p_{hi}] &=& \int\mbox{d}\vec x\, \left\{Q^\dagger
[\vec\alpha\cdot (-i\nabla - g\vec A^at^a) +\beta m](-i\vec\nabla-g\vec A)Q
\right.\nonumber\\
& &\left. -
Q^\dagger(-i\vec\nabla-g\vec A)[\vec\alpha\cdot (-i\nabla - g\vec A^at^a)
+\beta m]Q\right\}\nonumber\\
&=& \int\mbox{d}\vec x\, \left\{Q^\dagger (-i\stackrel{\leftarrow}{D_0})
(-i\vec\nabla-g\vec A)Q - Q^\dagger(-i\vec\nabla-g\vec A)(iD_0)Q\right\}\\
&=& \int\mbox{d}\vec x\, \left\{Q^\dagger [iD_0,\, (-i\vec\nabla-g\vec A)]
Q - i\partial_0[Q^\dagger (-i\vec\nabla-g\vec A)Q]\right\}\nonumber\\
&=& \int\mbox{d}\vec x\, Q^\dagger igF^a_{0i}t^aQ - i\frac{\mbox{d}p_{hi}}
{\mbox{d}t}\nonumber\,.
\end{eqnarray}
In the second equality the equation of motion for the field $Q$
(\ref{eqmotion}) has been used. The field tensor is defined as
$F_{\mu\nu}^a=\partial_\mu A_\nu^a-\partial_\nu A_\mu^a-gf_{abc}A_\mu^b
A_\nu^c$\,. The time-derivative on the right-hand side can be computed
with the help of the equation of motion for Heisenberg operators
$i\mbox{d}A/\mbox{d}t=[A,\,H]$, with $H$ the total hamiltonian (\ref{Ht}).
In this way one obtains the commutator between the brown muck hamiltonian
and the heavy quark momentum
\begin{equation}\label{intermediate}
[H,\, p_{hi}] = -\int\mbox{d}\vec x\, Q^\dagger igF^a_{0i}t^aQ\,.
\end{equation}

   So far everything has been completely general and we have not made use
of the fact that the quark $Q$ is heavy. The important simplification which
appears in the heavy mass limit is that it makes sense to speak about
the eigenstates of the brown muck in a hadron moving with a well-defined
momentum \cite{1}. This means that in this limit the brown muck hamiltonian
commutes with the total momentum:
\begin{equation}
[H,\, \vec p_t] = [H,\,\vec p\,] + [H,\,\vec p_h] = 0\,.
\end{equation}
{}From this relation and (\ref{intermediate}), one immediately obtains the
needed commutator between the brown muck operators
\begin{equation}\label{basic}
[H,\, p_i] = \int\mbox{d}\vec x\, Q^\dagger igF^a_{0i}t^aQ\,.
\end{equation}

   It is now a simple matter to compute $\mu_\chi^3$ (\ref{muchi}). It can
be written as a double commutator
\begin{equation}\label{double}
\mu_\chi^3 = \frac12 \langle 0|[p_i,\, [H,\, p_i]]|0\rangle\,,
\end{equation}
where the relation
\begin{equation}
\langle 0|[H,\vec p\,^2]|0\rangle = \langle 0|p_i[H,\,p_i]|0
\rangle + \langle 0|[H,\,p_i]p_i|0\rangle = 0
\end{equation}
has been used. The vanishing of this matrix element is due to the fact that
the states are eigenstates of $H$.
Further, note that the first $p_i$ can be replaced in (\ref{double}) by
$-p_{hi}$ since the external state (the heavy hadron is assumed at rest) is
an eigenstate of the total momentum
\begin{equation}
\vec p_t|0\rangle = 0\,.
\end{equation}
Thus one obtains
\begin{eqnarray}
\mu_\chi^3 &=& -\frac12 \langle 0|\,[p_{hi},\, [H,\, p_i]]|0\rangle\nonumber
\\&=&
-\frac12 \langle 0|\,[\int\mbox{d}^3x Q^\dagger(-i\partial_i + gA^a_i t^a)
Q,\, \int\mbox{d}^3y\,Q^\dagger igF_{0i}^a t^aQ]|0\rangle\nonumber\\
&=& \frac12 \langle 0|\int\mbox{d}^3x Q^\dagger gD^\mu F_{0\mu}^a t^aQ
|0\rangle \nonumber\\&=& \frac12 \langle 0|\int\mbox{d}^3x \bar Q\gamma^0
g(D^\mu F_{0\mu}^a)t^aQ|0\rangle\,,
\end{eqnarray}
which is the result quoted in the text (\ref{final}).

\end{document}